\documentclass[AMA,STIX1COL,]{WileyNJD-v2}

\usepackage{tikz}
\usepackage{pgfplots}
\usepackage{pdflscape}
\usepackage{color}
\usepackage{soul}
\usepackage{multirow}
\usepackage{multicol}
\usepackage{colortbl}
\usepackage{hhline}
\usepackage{longtable}
\usepackage{array}
\usepackage{hyperref}

\articletype{Research article}

\received{2021-01-01}

\revised{2021-02-01}

\accepted{2021-03-01}

\begin{document}

\title{Spike-and-Slab LASSO Generalized Additive Models and Scalable
Algorithms for High-Dimensional Data Analysis}

\author[a]{Boyi Guo*}
\author[b]{Byron C. Jaeger}
\author[a]{AKM Fazlur Rahman}
\author[a]{D. Leann Long}
\author[a]{Nengjun Yi*}

\authormark{Guo et al.}

\address[a]{Department of Biostatistics, University of Alabama at
Birmingham, Birmingham, USA}
\address[b]{Department of Biostatistics and Data Science, Wake Forest
School of Medicine, Winston-Salem, USA}

\corres{Boyi Guo and Nengjun Yi, Department of Biostatistics, University
of Alabama at Birmingham, Birmingham, USA. \email{boyiguo1@uab.edu},
\email{nyi@uab.edu}}

\presentaddress{This is sample for present address text this is sample
for present address text}

\abstract{There are proposals that extend the classical generalized
additive models (GAMs) to accommodate high-dimensional data (\(p>>n\))
using group sparse regularization. However, the sparse regularization
may induce excess shrinkage when estimating smooth functions, damaging
predictive performance. Moreover, most of these GAMs consider an
``all-in-all-out'' approach for functional selection, rendering them
difficult to answer if nonlinear effects are necessary. While some
Bayesian models can address these shortcomings, using Markov chain Monte
Carlo algorithms for model fitting creates a new challenge, scalability.
Hence, we propose Bayesian hierarchical generalized additive models as a
solution: we consider the smoothing penalty for proper shrinkage of
curve interpolation via reparameterization. A novel two-part
spike-and-slab LASSO prior for smooth functions is developed to address
the sparsity of signals while providing extra flexibility to select the
linear or nonlinear components of smooth functions. A scalable and
deterministic algorithm, EM-Coordinate Descent, is implemented in an
open-source R package BHAM. Simulation studies and metabolomics data
analyses demonstrate improved predictive and computational performance
against state-of-the-art models. Functional selection performance
suggests trade-offs exist regarding the effect hierarchy assumption.}


\maketitle

\pgfplotsset{compat=1.17}
\usetikzlibrary{shapes.geometric, arrows, positioning, calc, matrix, backgrounds, fit}
\newcommand{\bs}[1]{\boldsymbol{#1}}
\newcommand{\tp}{*}
\newcommand{\pr}{\text{Pr}}
\newcommand{\repa}{\text{repa}}
\newcommand{\simiid}{\overset{\text{iid}}{\sim}}

\section{Introduction}
\label{sec:intro}

Much modern biomedical research, e.g., sequencing data analysis and
electronic health record data analysis, require special treatment of
high-dimensionality, commonly known as \(p >> n\) problem. There is
extensive literature on high-dimensional linear models via penalized
models or Bayesian hierarchical models, see Mallick and Yi
\citep{Mallick2013} for review. These models are built upon a
restrictive and unrealistic assumption, linearity. In classical
statistical modeling, many strategies and models are proposed to relax
the linearity assumption with various degrees of complexity. For
example, variable categorization is a simple and common practice in
epidemiology but suffers from power and interpretation issues. More
complex models to address nonlinear effects include random forest and
other so-called ``black box'' models \citep{Breiman2001}. These models
are useful for statistical prediction but do not estimate parameters
relevant to the data generation process that one can draw inferences
from. In addition, how to generalize these ``black box'' models to the
high-dimensional setting remains unclear.

For their straightforward interpretation and flexibility, nonparametric
regression models serve as great alternatives to the ``black-box''
models in prediction and variable selection. Among those, generalized
additive models (GAMs), proposed in the seminal work of Hastie and
Tibshirani \citep{Hastie1987}, grew to be one of the most popular
modeling tools. In a GAM, the response variable, which is assumed to
follow some exponential family distribution, can be modeled with the
summation of smooth functions. Nevertheless, the classical GAMs cannot
fulfill the increasing analytic demands for high-dimensional data
analysis.

There exist some proposals to generalize the classical GAM to
accommodate high-dimensional applications. The regularized models,
branching out from group regularized linear models, are used to fit GAMs
by accounting for the structure introduced when expanding smooth
functions. Ravikumar et al. \citep{Ravikumar2009} extended the group
LASSO \citep{Yuan2006} to additive models (AMs); Huang et al.
\citep{Huang2010} further developed adaptive group LASSO for additive
models; Wang et al. \citep{Wang2007} and Xue \citep{Xue2009}
respectively applied group SCAD penalty \citep{Fan2001} to additive
models. Bayesian hierarchical models are also used in the context of
high-dimensional additive models, particularly within the spike-and-slab
literature. Various group spike-and-slab priors \citep{Xu2015, Yang2020}
combining with computationally intensive Markov chain Monte Carlo (MCMC)
algorithms are proposed, where the application on AMs is treated as a
special case. Bai and co-authors\citep{Bai2020} were the first to apply
group spike-and-slab LASSO prior to Gaussian AMs using a fast
optimization algorithm and further generalized the framework to GAMs
\citep{Bai2021}. Focus on addressing the sparsity, these methods can
excessively penalize the bases of a smooth function and produce
inaccurate predictions, particularly when complex signals are assumed
and large numbers of knots are used. \citep{Scheipl2013} In addition,
these methods adopt an `all-in-all-out' strategy, i.e.~either including
or excluding the variable completely, rendering no space for bi-level
selection. Scheipl et al. \citep{Scheipl2012} proposed a spike-and-slab
structure prior that addresses the bi-level selection. But the model
fitting relies on computationally intensive MCMC algorithms and creates
scalability concerns. Developing a fast, flexible and accurate
generalized additive model framework would be of special interest.

We propose a novel Bayesian hierarchical generalized additive model
(BHAM) for outcome prediction in the context of high-dimensional data
analysis. Specifically, we incorporate smoothing penalties, derived from
the smoothing spline literature \cite{Wood2017}, via reparameterization
of smooth functions to avoid excessive shrinkage on the bases. Smoothing
penalties were also previously used in the spike-and-slab GAM
\cite{Scheipl2012} and the sparsity-smoothness penalty \cite{Meier2009}.
We then impose a new two-part spike-and-slab LASSO prior to address the
signal sparsity. In addition, a scalable optimization-based algorithm,
EM-Coordinate Descent (EM-CD) algorithm, is developed. While the primary
focus of this model is to improve prediction, the proposed model also
provides utility in functional selection. Notably, the two-part prior
that follows the effect hierarchy principle motivates a bi-level
selection, rendering one of three possibilities for each predictor: no
effect, only linear effect, or linear and nonlinear effects. The
proposed model is implemented in a publicly available R package
\texttt{BHAM} via \url{https://github.com/boyiguo1/BHAM}.

The proposed framework, BHAM, differs from previous spike-and-slab based
GAMs, i.e., the spike-and-slab GAM \citep{Scheipl2012} and the sparse
Bayesian GAM (SB-GAM) \citep{Bai2021}, in three ways. Firstly, the
proposed prior for smooth functions is a spike-and-slab LASSO type prior
using independent mixture double exponential distribution, compared to
spike-and-slab GAM that uses normal-mixture-of-inverse gamma prior.
Spike-and-slab LASSO priors provide computational convenience during
model fitting by using optimization algorithms instead of intensive
sampling algorithms. They make fitting high-dimensional models more
feasible without sacrificing performance in prediction and variable
selection. Secondly, SB-GAM uses a group spike-and-slab LASSO prior with
an EM-CD algorithm to fit the model. While both methods use the
combination of expectation maximization algorithm and coordinate descent
algorithm, there are subtle differences in the implementation due to the
difference in prior specification. The proposed model sets up
independent priors among basis coefficients after the reparameterization
step, which provides some advantage in computation. Lastly, the proposed
model addresses the incapability of bi-level selection in SB-GAM.

In Section \ref{sec:BHAM}, we establish the Bayesian hierarchical
generalized additive model, introduce the proposed spike-and-slab spline
priors, and describe the fast-fitting EM-CD algorithm. In Section
\ref{sec:sim}, we compare the proposed framework to state-of-the-art
models via Monte Carlo simulation studies. Analyses of two metabolomics
datasets are presented in Section \ref{sec:real_data}. Conclusion and
discussions are given in Section \ref{sec:concl}.

\section{Bayesian Hierarchical Additive Models (BHAM)}
\label{sec:BHAM}

We assume the response variable, \(Y\), follows an exponential family
distribution with density function \(f(y)\), mean \(\mu\) and dispersion
parameter \(\phi\). The mean of the response variable can be modeled as
the summation of smooth functions, \(B_j(\cdot), j = 1, \dots, p\), of a
given \(p\)-dimensional vector of predictors \(\boldsymbol{x}\), written
as \begin{equation}\label{eq:gam}
 E(Y|\boldsymbol{x}) = g^{-1}(\beta_0 + \sum\limits^p_{j=1}B_j(x_j)) = g^{-1}(\beta_0 + \sum\limits^p_{j=1} \boldsymbol{\beta}_j^T \boldsymbol{X}_j),
\end{equation} where \(g^{-1}(\cdot)\) is the inverse of a monotonic
link function. Given \(n\) data points
\(\{y_i, \boldsymbol{x}_i\}^n_{i=1}\), the data distribution is
expressed as \begin{equation}
f(\boldsymbol{Y} = \boldsymbol{y}| \boldsymbol{\beta}, \phi) = \prod\limits^n_{i=1}f( Y = y_i|\boldsymbol{\beta}, \phi).\nonumber
\end{equation}

The basis function matrix, i.e.~the design matrix derived from the
smooth function \(B_j(x_j)\), is denoted \(\boldsymbol{X}_j\) for the
variable \(x_j\). The dimension of the design matrix depends on the
choice of the smooth function, and is denoted as \(K_j\) for \(x_j\).
\(\boldsymbol{\beta}_j\) denotes the basis function coefficients for the
\(j\)th variable such that
\(B_j(x_j) = \boldsymbol{\beta}_j^T \boldsymbol{X}_j\). With slight
abuse of notation, we denote vectors and matrices in bold fonts
\(\boldsymbol{\beta}, \boldsymbol{X}\) with conformable dimensions,
where scalar and random variables are denoted in unbold fonts
\(\beta, X\). The matrix transposing operation is denoted with a
superscript \(^T\). To note, the proposed model can include parametric
forms of variables in the model, and hence considers general linear
models and semiparametric regression models as special cases.

\subsection{Smooth Function Reparameterization}

To encourage proper smoothing of each additive function, we adopt the
smoothing penalty from smoothing spline models\citep{Wood2017}. A
smoothing penalty is the quadratic norm of the basis coefficients and
allows different shrinkage on different bases, mathematically
\begin{equation}
  \text{pen}\left[B_j(x)\right] = \lambda_j \int B^{\prime\prime}_j(x)^2dx = \lambda_j \boldsymbol{\beta}_j^T \boldsymbol{S}_j \boldsymbol{\beta}_j ,\nonumber
\end{equation} where \(\boldsymbol{S}_j\) is a known smoothing penalty
matrix and \(\lambda_j\) denotes a smoothing parameter. A linear
function can be modeled as \(B_j(x_j) = x_j\) with the smoothing penalty
matrix \(\boldsymbol{S}_j = \begin{bmatrix}0\end{bmatrix}\). Unlike
previous regularized methods that either ignore the smoothing penalty
completely or restrain the smoothing penalty as a component of sparse
penalty which leads to a more restrictive solution, we consider an
additional mechanism in pair with the proposed prior (described in
Section \ref{sec:method_prior}) to address the smoothness and sparsity
in signals such that the locally adaptive nature of the smoothing
penalty retains.

Marra and Wood \citep{Marra2011} proposed a reparameterization procedure
to factor the smoothing penalty into the design matrix of each smooth
function. Given that the smoothing penalty matrix \(\boldsymbol{S}_j\)
is symmetric and positive semi-definite for the univariate smooth
functions, we eigendecompose the penalty matrix
\(\boldsymbol{S} = \boldsymbol{U} \boldsymbol{D} \boldsymbol{U}^T\) ,
where the matrix \(\boldsymbol{D}\) is diagonal with the eigenvalues
arranged in the ascending order. To note, \(\boldsymbol{D}\) can contain
elements of zeros on the diagonal, where the zeros are associated with
the linear space of the smooth function. For the most popular smooth
function, cubic splines, the dimension of the linear space is one.
Hereafter, we focus on discussing a uni-dimensional linear space for
simplicity; however, it generalizes easily to the cases where the linear
space is multidimensional. We further write the orthonormal matrix
\(\boldsymbol{U} \equiv \begin{bmatrix} \boldsymbol{U}^0 : \boldsymbol{U}^{*}\end{bmatrix}\)
containing the eigenvectors as columns in the corresponding order to
\(\boldsymbol{D}\). That is, \(\boldsymbol{U}\) contains the
eigenvectors \(U^0\) with zero eigenvalues for the linear space and
\(\boldsymbol{U}^{*}\) contains the eigenvectors (as columns) for the
non-zero eigenvalues, i.e.~the nonlinear space. We multiply the basis
function matrix \(\boldsymbol{X}\) with the orthonormal matrix
\(\boldsymbol{U}\) for the new design matrix
\({\boldsymbol{X}}^\text{repa}= \boldsymbol{X} \boldsymbol{U} \equiv \begin{bmatrix} X^0 : \boldsymbol{X}^{*} \end{bmatrix}\).
An additional scaling step is imposed on \(\boldsymbol{X}^{*}\) by the
non-zero eigenvalues of \(\boldsymbol{D}\) such that the new basis
function matrix \(\boldsymbol{X}^\ast\) can receive a uniform penalty on
each of its dimensions. With slight abuse of the notation, we drop the
superscript \(^\text{repa}\) and denote
\(\boldsymbol{X}_j \equiv \begin{bmatrix} X_j^0 : \boldsymbol{X}_j^{*} \end{bmatrix}\)
as the basis function matrix for the \(j\)th variable after the
reparameterization. A spline function can be expressed in the matrix
form \[
B_j(x_j) = B_j^0(x_j) + B_j^*(x_j) = \beta_j X^0_j + \boldsymbol{\beta_j^*}^T \boldsymbol{X}_j^*,
\] and the generalized additive model in Equation (\ref{eq:gam}) now is
\begin{equation}\label{eq:gam-repa}
E(Y|\boldsymbol{x}) = g^{-1}(\beta_0 + \sum\limits^p_{j=1} B_j(x_j)) = g^{-1}(\beta_0 + \sum\limits^p_{j=1} \boldsymbol{\beta}_j^T \boldsymbol{X}_j) = g^{-1}\left[\beta_0 + \sum\limits^p_{j=1} (\beta_j X^0_j + {\boldsymbol{\beta}_j^*}^T \boldsymbol{X}_j^*)\right],
\end{equation} where the coefficients
\(\boldsymbol{\beta}_j \equiv \begin{bmatrix} \beta_j : \boldsymbol{\beta}^*_j \end{bmatrix}\)
is an augmentation of the coefficient scalar \(\beta_j\) of linear space
and the coefficient vector \(\boldsymbol{\beta}^*_j\) of nonlinear
space.

To summarize, the reparameterization step provides three benefits.
Firstly, the reparameterization integrates the smoothing penalty matrix
into the design matrix, and encourages models to properly smooth the
nonlinear function when sparsity penalty exists. Secondly, the
eigendecomposition of the smoothing penalty matrix allows the isolation
of the linear space from the nonlinear space, improving the feasibility
of bi-level functional selection. Lastly, the eigendecomposition
facilitates the construction of an orthonormal design matrix, which
makes imposing independent priors on the coefficients possible. This
reduces the computational complexity compared to using a multivariate
priors, and improve the generalizability of the framework to be
compatible with other choices of priors.

\subsection{Two-part Spike-and-Slab LASSO Prior for Smooth Functions}\label{sec:method_prior}

The family of spike-and-slab regression models is one of the most
commonly used models in high-dimensional data analysis for its utility
in outcome prediction and variable selection. Among all the
spike-and-slab priors, the spike-and-slab LASSO (SSL) prior
\cite{Rockova2018b, Rockova2018} is one of the most popular choices
because it's highly scalable. The SSL prior is composed of two double
exponential distributions with mean 0 and different dispersion
parameters, \(0 < s_0 < s_1\), mathematically, \begin{equation} 
\beta | \gamma \sim (1-\gamma)DE(0, s_0) + \gamma DE(0, s_1), 0 < s_0 < s_1.\nonumber
\end{equation} The latent binary variable \(\gamma \in \{0,1\}\)
indicates whether a variable \(x\) is included in the model, while the
dispersion parameters \(s_0\) and \(s_1\) control the shrinkage of the
coefficient. Given that both double exponential distributions have a
mean of 0 and the latent indicator \(\gamma\) can only take the value of
0 or 1, the mixture double exponential distribution can be formulated as
one single double exponential density, \begin{equation} \label{eq:ssl}
\beta | \gamma \sim DE(0, (1-\gamma)s_0 + \gamma s_1), 0 < s_0 < s_1.
\end{equation} Compared to other priors for high-dimensional data
analysis, SSL has the following advantages. First of all, the SSL prior
provides a locally adaptive shrinkage when estimating the coefficients.
Secondly, the SSL prior encourages a sparse solution, making variable
selection straightforward. Thirdly, the SSL prior motivates a scalable
algorithm, the EM-CD algorithm, for model fitting, and hence is more
feasible for high-dimensional data analysis. We defer to Bai et
al.~\cite{Bai2021Review} for a detailed discussion.

We introduce a novel SSL-based prior for smooth functions in GAMs. Given
the reparameterized design matrix
\(\boldsymbol{X}_j = \begin{bmatrix} X^0_j : \boldsymbol{X}_j^*\end{bmatrix}\)
for the \(j\)th variable, we impose a two-part SSL prior to the
coefficients
\(\boldsymbol{\beta}_j = \begin{bmatrix} \beta_j : \boldsymbol{\beta}_j^*\end{bmatrix}\).
Specifically, the linear space coefficient has an SSL prior and the
nonlinear space coefficients share a group SSL prior,
\begin{align}\label{eq:bham_ssl}
  \beta_{j} | \gamma_{j},s_0,s_1 &\sim DE(0,(1-\gamma_{j}) s_0 + \gamma_{j} s_1) \nonumber \\
  \beta^*_{jk} | \gamma^*_{j},s_0,s_1 &\overset{\text{iid}}{\sim}DE(0,(1-\gamma^*_{j}) s_0 + \gamma^*_{j} s_1), k=1,\dots, K_j
\end{align} where \(\gamma_{j}\in\{0,1\}\) and
\(\gamma^*_{j}\in \{0,1\}\) are two latent indicator variables,
indicating if the model includes the linear effect and the nonlinear
effect of the \(j\)th variable respectively. \(s_0\) and \(s_1\) are
scale parameters, assuming \(0 < s_0 < s_1\) and given. These scale
parameters \(s_0\) and \(s_1\) can be treated as tuning parameters and
optimized via cross-validation, discussed in Section \ref{sec:tune}.

The proposed two-part SSL prior, particularly the group SSL prior of the
nonlinear space coefficients, differs from previous group SSL priors
\cite{Tang2018, Tang2019}, as the proposed prior follows the effect
hierarchy principle. Effect hierarchy refers to the principle that
``lower-order effects are more likely to be active than higher-order
effects'' defined by Chipman\cite{chipman2006prior}. To implement, we
consider the shared latent indicator of nonlinear coefficients
\(\gamma^*_j\) depends on the value of the linear space latent indicator
\(\gamma_j\), while both latent indicators \(\gamma_j\) and
\(\gamma^*_j\) follow a Bernoulli distribution. While the probability of
including the linear effect is \(\theta_j\), the probability of
including the nonlinear effect is \(\gamma_{j}\theta_j\). \[
\begin{aligned}
&\gamma_{j} | \theta_j \sim Bin(1, \theta_j) & & 
&\gamma_{j}^*| \gamma_{j}, \theta_j \sim Bin(1, \gamma_{j}\theta_j).
\end{aligned}
\] This is, when the linear effect is not selected, the probability of
including the nonlinear effect drops from \(\theta_j\) to 0. For
computational convenience, we analytically integrate \(\gamma_j\) out
such that \(\gamma_{j}^*| \theta_j \sim Bin(1, \theta_j^2)\) (see the
derivation in the Supporting Information).

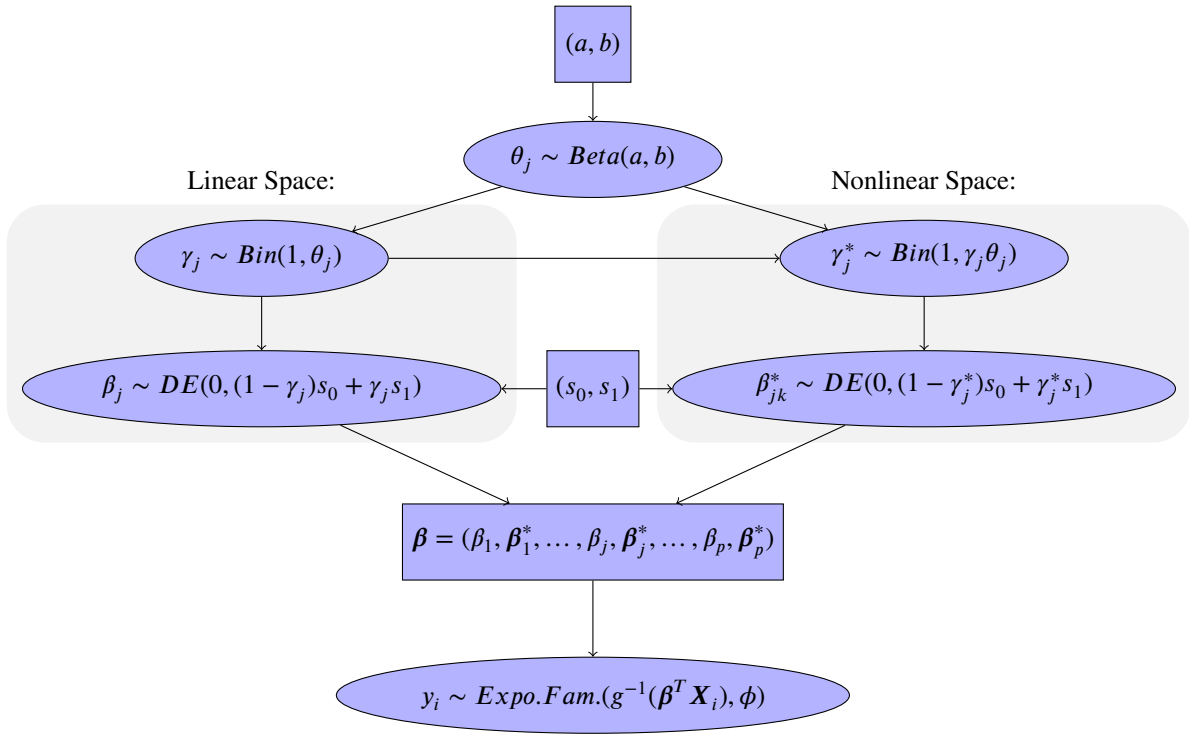
\begin{figure}
\centering
\begin{tikzpicture} [
staticCompo/.style = {rectangle, minimum width=1cm, minimum height=1cm,text centered, draw=black, fill=blue!30},
outCome/.style={ellipse, minimum width=3cm, minimum height=1cm,text centered, draw=black, fill=blue!30},
mymatrix/.style={matrix of nodes, nodes=outCome, row sep=1em},
PriorBoarder/.style={rectangle, minimum width=5cm, minimum height=10cm, text centered, fill=lightgray!30},
background/.style={rectangle, fill=gray!10,inner sep=0.2cm, rounded corners=5mm}
]

\matrix (linearPrior) [matrix of nodes, column sep = 0mm, row sep = 0.7cm] {
  \node (linearGamma) [outCome] { $\gamma_j \sim Bin(1, \theta_j) $ };\\
  \node (linearBeta) [outCome] { $\beta_j \sim DE(0,(1-\gamma_{j}) s_0 + \gamma_{j} s_1)$};\\
};
\matrix (penPrior) [right = 2cm of linearPrior, matrix of nodes, column sep = 0mm, row sep = 0.7cm] {
  \node (penGamma) [outCome] { $\gamma_{j}^\tp \sim Bin(1, \gamma_{j}\theta_j)$ };\\
  \node (penBeta) [outCome] { $\beta_{jk}^\tp \sim  DE(0,(1-\gamma^\tp_{j}) s_0 + \gamma^\tp_{j} s_1)$};\\
};

\node (s) [staticCompo]  at ($(linearBeta)!0.5!(penBeta)$)  {($s_0, s_1$)};
\node (Beta) [staticCompo, below = 1cm of s] {$\bs \beta = (\beta_1, \bs \beta^\tp_1, \dots,\beta_j, \bs \beta^\tp_j , \dots,\beta_p, \bs \beta^\tp_p) $};
\node (Theta)[outCome, above = 2cm of s] {$\theta_{j} \sim Beta(a, b)$};
\node (ab)[staticCompo, above = 0.5cm of Theta] {$(a, b)$};
\node (Y) [outCome, below = 1cm of Beta] {$y_i \sim Expo. Fam. (g^{-1}(\bs \beta^T \bs X_i), \phi)$};

\draw[->] (Theta) -- (linearGamma);
\draw[->] (Theta) -- (penGamma);
\draw[->] (linearGamma) -- (linearBeta) ;
\draw[->] (penGamma) -- (penBeta);
\draw[->] (linearGamma) -- (penGamma);
\draw[->] (ab) -- (Theta);
\draw[->] (s) -- (linearBeta) ;
\draw[->] (s) -- (penBeta);
\draw[->] (linearBeta) -- (Beta);
\draw[->] (penBeta) -- (Beta);
\draw[->] (Beta) --  (Y);

\begin{pgfonlayer}{background}
  \node [background,
   fit=(linearGamma) (linearBeta),
   label=above:Linear Space:] {};
  \node [background,
    fit=(penGamma) (penBeta),
    label=above:Nonlinear Space:] {};
\end{pgfonlayer}

\end{tikzpicture}

\caption{Directed acyclic graph of the proposed Bayesian hierarchical additive model with parameter expansion. Elliposes are stochastic nodes, rectangles and are deterministic nodes. }
\label{fig:SSprior}
\end{figure}

To allow the shrinkage to self-adapt to the sparsity and smoothing
pattern of the data, we further specify the parameter \(\theta_j\)
follows a beta distribution with given shape parameters \(a\) and \(b\),
\[
\theta_j \sim Beta(a, b).
\] The beta distribution is a conjugate prior to the binomial
distribution and hence provides some computation convenience. Having a
prior distribution of \(\theta_j\) enables the proposed prior to inherit
the selective shrinkage property and self-adaptivity
\cite{Bai2021Review} from the classical SSL prior. In other words, when
a smooth function is significant, the coefficients of the smooth
function escape the overall shrinkage and produce a more accurate
estimate, particularly in pair with the smoothing penalty implicitly
addressed via the reparameterization. Meanwhile, the hyper prior
encourages information borrowing across coordinates and hence automatic
adjust for different levels of sparsity. Hereafter, we refer to the
Bayesian hierarchical generalized additive models with the two-part
spike-and-slab LASSO prior as BHAM, and visually presented in Figure
\ref{fig:SSprior}.

\subsection{Scalable EM-Coordinate Descent Algorithm}

Despite the advantage to estimate posterior densities, using MCMC
algorithms to fit the proposed model is computationaly prohibited and
not feasible for high-dimensional data. Previous research shows the
computation performance of MCMC algorithms for spike-and-slab models is
bottlenecked for medium-sized data (\(p\)=25) \cite{George1997}, and
substantially slows as \(p\) increases modestly in the GAM context
\cite{Scheipl2013}. Hence, we consider the optimization algorithms that
focus on the maximum a posteriori estimates at the cost of posterior
inference. Specifically, we extend the EM-Coordinate Descent (EM-CD)
algorithm to fit BHAMs. Similar to the EMVS algorithm
\cite{Rockova2014a} for spike-and-slab models, the EM-CD algorithm is
based on the expectation-maximization (EM) algorithm, integrating the
Coordinate Descent algorithm in each iterative step to find the
posterior mode. The EM-CD algorithm has been well adapted in generalized
linear models \cite{Tang2017a}, Cox proportional hazards models
\cite{Tang2017}, and their grouped counterparts
\cite{Tang2018, Tang2019}. The EM-CD algorithm provides deterministic
solutions, which becomes a popular property for reproducible research.

For BHAMs, we define the parameters of interest as
\(\Theta = \{\boldsymbol{\beta}, \boldsymbol{\theta}, \phi\}\) and
consider the latent binary indicators \(\boldsymbol{\gamma}\) as
nuisance parameters of the model, in other words, the ``missing'' data
in the EM context. Our objective is to find the parameters \(\Theta\)
that maximize the posterior density function, or equivalently the
logarithm of the density function, \[
\begin{aligned}
& \text{argmax}_{\Theta}
\log f(\Theta, \boldsymbol{\gamma}| \textbf{y}, \textbf{X}) \\
&= \log f(\textbf{y}|\boldsymbol{\beta}, \phi) + \sum\limits_{j=1}^p\left[\log f(\beta_j|\gamma_j)+\sum\limits_{k=1}^{K_j} \log f(\beta^{*}_{jk}|\gamma^{*}_{j})\right]\\
& +\sum\limits_{j=1}^{p} \left[ (\gamma_j+\gamma_{j}^{*})\log \theta_j + (2-\gamma_j-\gamma_{j}^{*}) \log (1-\theta_j)\right] +  \sum\limits_{j=1}^{p}\log f(\theta_j),
\end{aligned}
\]\\
where \(f(\textbf{y}|\boldsymbol{\beta}, \phi)\) is the data
distribution and \(f(\theta)\) is the Beta(a, b) density. We choose
non-informative prior for the intercept \(\beta_0\) and the dispersion
parameter \(\phi\); for example, \(f(\beta_0|\tau_0^2)=N(0,\tau_0^2)\)
with \(\tau^2_0\) set to a large value and \(f(\log \phi) \propto 1\).

We use the EM algorithm to find the maximum a posteriori estimate of
\(\Theta\). This is, in the E-step, we calculate the expectation of
posterior density function of
\(\log f(\Theta, \boldsymbol{\gamma}| \textbf{y}, \textbf{X})\) with
respect to the latent indicators \(\boldsymbol{\gamma}\) conditioning on
the parameter values from previous iteration \(\Theta^{(t-1)}\), \[
E_{\boldsymbol{\gamma}|\Theta^{(t-1)}}\log f(\Theta, \boldsymbol{\gamma}| \textbf{y}, \textbf{X}) .
\] Hereafter, we use the shorthand notation
\(E(\cdot)\equiv E_{\boldsymbol{\gamma}|\Theta^{(t-1)}}(\cdot)\). In the
M-step, we find the parameters \(\Theta^{(t)}\) that maximize
\(E\log f(\Theta, \boldsymbol{\gamma}| \textbf{y}, \textbf{X})\). The
parenthesized subscription \(^{(t)}\) denotes the parameter estimation
at the \(t\)th iteration. The E- and M- steps are iterated until the
algorithm converges.

To note here, the log-posterior density of BHAMs (up to additive
constants) can be written as a two-part equation
\[ \log f(\Theta, \boldsymbol{\gamma}| \textbf{y}, \textbf{X}) = Q_1(\boldsymbol{\beta}, \phi) + Q_2 (\boldsymbol{\gamma},\boldsymbol{\theta}),\]
where
\[ Q_1\equiv Q_1(\boldsymbol{\beta}, \phi) = \log f(\textbf{y}|\boldsymbol{\beta}, \phi) + \sum\limits_{j=1}^p\left[\log f(\beta_j|\gamma_j)+\sum\limits_{k=1}^{K_j} \log f(\beta^{*}_{jk}|\gamma^{*}_{jk})\right]\]
and
\[Q_2 \equiv Q_2(\boldsymbol{\gamma},\boldsymbol{\theta}) = \sum\limits_{j=1}^{p} \left[ (\gamma_j+\gamma_{j}^{*})\log \theta_j + (2-\gamma_j-\gamma_{j}^{*}) \log (1-\theta_j)\right] +  \sum\limits_{j=1}^{p}\log f(\theta_j).\]
\(Q_1\) and \(Q_2\) are respectively the log posterior density of the
coefficients \(\boldsymbol{\beta}\) and the log posterior density of the
probability parameters \(\boldsymbol{\theta}\) conditioning on
\(\boldsymbol{\gamma}\). Meanwhile, conditioning on
\(\boldsymbol{\gamma}\), \(Q_1\) and \(Q_2\) are independent and can be
maximized separately for \(\boldsymbol{\beta}, \phi\) and
\(\boldsymbol{\theta}\). With the proposed two-part spike-and-slab LASSO
prior, \(Q_1\) can be treated as penalized likelihood function and
maximization of \(E(Q_1)\) can be solved via the Coordinate Descent
algorithm in each iteration. Coordinate descent is an optimization
algorithm that offers extreme computational advantages, and is famous
for its application in optimizing the \(l_1\) penalized likelihood
function. Maximization of \(E(Q_2)\) can be solved via closed-form
equations following the beta-binomial conjugate relationship.

The density function of the mixture double exponential prior of
coefficient \(\beta\) can be written as \[
f(\beta|\gamma, s_0, s_1) = \frac{1}{2\left[(1-\gamma)s_0 + \gamma s_1\right]}\exp(-\frac{|\beta|}{(1-\gamma)s_0 + \gamma s_1}),
\] and \(E(Q_1)\) can be expressed as a log-likelihood function with
\(l_1\) penalty \begin{equation}\label{eq:Q1_CD}
E(Q_1) = \log f(\textbf{y}|\boldsymbol{\beta}, \phi) - \sum\limits_{j=1}^p\left[E({S_j}^{-1})|\beta_j|+\sum\limits_{k=1}^{K_j}E({S^{*}}^{-1}_{j})|\beta_{jk}|\right],
\end{equation} where \(S_{j} = (1-\gamma_{j}) s_0 + \gamma_{j} s_1\) and
\(S^*_{j} = (1-\gamma^*_{j}) s_0 + \gamma^*_{j} s_1\). To calculate two
unknown quantities \(E({S_j}^{-1})\) and \(E({S^*}^{-1}_j)\), the
posterior probability
\(p_{j} \equiv \text{Pr}(\gamma_{j}=1|\Theta^{(t-1)})\) and
\(p_{j}^*\equiv \text{Pr}(\gamma^*_{j}=1|\Theta^{(t-1)})\) are
necessary, which can be derived via Bayes' theorem. The calculation of
\(p_j^*\) is slightly different from that of \(p_j\), as \(p_j^*\)
depends on the values of the vector \(\boldsymbol{\beta}^*_{j}\) and
\(p_j\) only depends on the scalar \(\beta_j\). The calculation follows
the equations below, \begin{align*}
p_{j} &= \frac{\text{Pr}(\gamma_{j} = 1|\theta_j)f(\beta_{j}|\gamma_{j}=1, s_1) }{\text{Pr}(\gamma_{j} = 1|\theta_j)f(\beta_{j}|\gamma_{j}=1, s_1) + \text{Pr}(\gamma_{j} = 0|\theta_j)f(\beta_{j}|\gamma_{j}=0, s_0)}\\
p^*_{j} &= \frac{\text{Pr}(\gamma^{*}_{j} = 1|\theta_j)\prod\limits_{k=1}^{K_j}f(\beta_{jk}|\gamma^{*}_{j}=1, s_1) }{\text{Pr}(\gamma^{*}_{j} = 1|\theta_j)\prod\limits_{k=1}^{K_j}f(\beta_{jk}|\gamma^{*}_{j}=1, s_1) + \text{Pr}(\gamma^{*}_{j} = 0|\theta_j)\prod\limits_{k=1}^{K_j}f(\beta_{jk}|\gamma^{*}_{j}=0, s_0)}
\end{align*} where \(\text{Pr}(\gamma_{j} = 1|\theta_j) = \theta_j\),
\(\text{Pr}(\gamma_{j} = 0|\theta_j) = 1-\theta_j\),
\(\text{Pr}(\gamma_{j}^*= 1|\theta_j) = \theta_j^2\),
\(\text{Pr}(\gamma_{j}^*= 0|\theta_j) = 1-\theta^2_j\),
\(f(\beta|\gamma=1, s_1) = \text{DE}(\beta|0 , s_1)\),
\(f(\beta|\gamma=0, s_0) = \text{DE}(\beta|0 , s_0)\). It is trivial to
show \begin{align*}\label{eq:exp_scale}
&E(\gamma_{j})  = p_{j} & &E(\gamma^{*}_{j}) = p_{j}^{*}\nonumber\\
&E({S}^{-1}_{j}) = \frac{1-p_{j}}{s_0} + \frac{p_{j}}{s_1} & &E({S_{j}^*}^{-1}) = \frac{1-p_{j}^{*}}{s_0} + \frac{p_{j}^{*}}{s_1}.
\end{align*} After replacing the calculated quantities, \(E(Q_1)\) can
be seen as a \(l_1\) penalized likelihood function with the
regularization parameter \(\lambda = E(S^{-1})\), and hence be optimized
via coordinate descent algorithm \citep{Friedman2010}. Independently,
the remaining parameters of interest \(\boldsymbol{\theta}\) can be
updated by maximizing \(E(Q_2)\). As the beta distribution is a
conjugate prior for Bernoulli distribution, \(\boldsymbol{\theta}\) can
be easily updated with a closed-form equation,
\begin{equation}\label{eq:update_theta}
\theta_j = \frac{p_j + p^*_{j} + a - 1 }{a + b}.
\end{equation}

Totally, the proposed EM-CD algorithm is summarized as follows:

\begin{enumerate}
\def\labelenumi{\arabic{enumi})}
\item
  Choose a starting value \(\boldsymbol{\beta}^{(0)}\) and
  \(\boldsymbol{\theta}^{(0)}\) for \(\boldsymbol{\beta}\) and
  \(\boldsymbol{\theta}\). For example, we can initialize
  \(\boldsymbol{\beta}^{(0)} = \boldsymbol{0}\) and
  \(\boldsymbol{\theta}^{(0)} = \boldsymbol{0}.5\)
\item
  Iterate over the E-step and M-step until convergence

  E-step: calculate \(E(\gamma_{j})\), \(E(\gamma^*_{j})\) and
  \(E({S}^{-1}_{j})\), \(E({S^*}^{-1}_{j})\) with estimates of
  \(\Theta^{(t-1)}\) from previous iteration

  M-step:

  \begin{enumerate}
  \def\labelenumii{\alph{enumii})}
  \item
    Update \(\boldsymbol{\beta}^{(t)}\), and the dispersion parameter
    \(\phi^{(t)}\) if exists, using the coordinate descent algorithm
    with the penalized likelihood function in Equation (\ref{eq:Q1_CD})
  \item
    Update \(\boldsymbol{\theta}^{(t)}\) using Equation
    (\ref{eq:update_theta})
  \end{enumerate}
\end{enumerate}

We assess convergence by the criterion:
\(|d^{(t)}-d^{(t-1)}|/(0.1+|d^{(t)}|)<\epsilon\), where
\(d^{(t)} = -2\log f(\textbf{y}| \textbf{X}, \boldsymbol{\beta}^{(t)},\phi^{(t)})\)
is the estimate of deviance at the \(t\)th iteration, and \(\epsilon\)
is a small value (say \(10^{-5}\)).

\subsection{Selecting Optimal Scale Values}
\label{sec:tune}

Our proposed model, BHAM, requires two preset scale parameters (\(s_0\),
\(s_1\)). Hence, we need to find the optimal values for the scale
parameters such that the model reaches its best prediction performance
regarding criteria of preference. This would be achieved by constructing
a two-dimensional grid, consisting of different (\(s_0\), \(s_1\))
pairs. However, previous research suggests the value of slab scale
\(s_1\) has less impact on the final model and is recommended to be set
as a generally large value, e.g.~\(s_1 = 1\), that provides no or weak
shrinkage. \citep{Rockova2018} As a result, we focus on examining
different values of spike scale \(s_0\). Instead of the two-dimensional
grid, we consider a sequence of \(L\) decreasing values
\(\{s_0^l\}: 0 < s_0^1 < s_0^2 < \dots < s_0^L < s_1\). Increasing the
spike scale \(s_0\) tends to include more non-zero coefficients in the
model. A measure of preference calculated with cross-validations (CV),
e.g.~deviance (defined as model log-likelihood times -2,
\(-2\log f(\boldsymbol{y}|\boldsymbol{\hat\beta}, \hat\phi)\)), area
under the curve (AUC), mean squared error, can be used to facilitate the
selection of a final model. The procedure is similar to the LASSO
implementation in the widely used R package \texttt{glmnet}, which
quickly fits LASSO models over a list of candidate values of \(\lambda\)
and gives a sequence of models for users to choose from.

\section{Simulation Study}
\label{sec:sim}

In this section, we compare the performance of the proposed model to six
alternative models: linear LASSO models, component selection and
smoothing operator (COSSO) \citep{Zhang2006GAM}, adaptive COSSO
\citep{Storlie2011}, generalized additive models with automatic
smoothing (referred to as \textit{mgcv} hereafter)\citep{Wood2011},
spike-and-slab GAM \cite{Scheipl2012}, and SB-GAM \citep{Bai2021}. We
use linear LASSO models as the benchmark, examining the performance when
the linearity assumption doesn't hold. COSSO is one of the earliest
smoothing spline models that consider sparsity-smoothness penalty.
Adaptive COSSO improved upon COSSO by using adaptive weight for
penalties such that the penalty of each functional component is
different for extra flexibility. \textit{mgcv} is one of the most
popular models for nonlinear effect interpolation and prediction.
Nevertheless, mgcv doesn't support analyses when the number of
parameters exceeds the sample size. Spike-and-slab GAM employs a
spike-and-slab prior for GAM and uses an MCMC algorithm for model
fitting. SB-GAM is the first spike-and-slab LASSO GAM. We implement
linear LASSO model with R package \texttt{glmnet 4.1-2}, COSSO and
adaptive COSSO with R package \texttt{cosso 2.1-1}, generalized additive
models with automatic smoothing with R package \texttt{mgcv 1.8-31},
spike-and-slab GAM with R package \texttt{spikeSlabGAM 1.1-15}, and
SB-GAM with R package \texttt{sparseGAM 1.0}. COSSO models and SB-GAM do
not provide flexibility to define smooth functions, and hence use the
default choices; mgcv, spikeSlabGAM and the proposed model allow
customized smooth functions and we choose the cubic regression spline.
We control the dimensionality of each smooth function, 10 bases, for all
different choices of smooth functions. We use a 5-fold CV with the
default selection criteria to select the final model for linear LASSO
model, COSSO models, SB-GAM and the proposed model. Twenty default
candidates of tuning parameters (\(s_0\) in BHAM, \(\lambda_0\) in
SB-GAM) are examined for SB-GAM and the proposed model that allow user
specification of tuning candidates. All computation was conducted on a
high-performance 64-bit Linux platform with 48 cores of 2.70GHz
eight-core Intel Xeon E5-2680 processors and 24G of RAM per core and R
3.6.2 \citep{R}.

Other related methods for high-dimensional GAMs also exist, notably the
methods of sparse additive models by Ravikumar et al.
\citep{Ravikumar2009}. However, we exclude these methods from the
current simulation study because they demonstrated inferior predictive
performance compared to \textit{mgcv} \citep{Scheipl2013}.

\subsection{Monte Carlo Simulation Study}

We follow the data generating process described in Bai \citep{Bai2021}:
we first generate \(n=500\) training data points with
\(p=4, 10, 50, 100, 200\) predictors respectively, where the predictors
\(\boldsymbol{X}\) are simulated from a multivariate normal distribution
\(\text{MVN}_{n\times p}(0, I_{P})\). We then simulate the outcome \(Y\)
from two distributions, Gaussian and binomial with the identity link and
logit link \(g(x) = \log(\frac{x}{1-x})\) respectively. The mean of each
outcome is derived via the following function \[
\mathbb{E}(Y) = g^{-1}(5 \sin(2\pi x_1) - 4 \cos(2\pi x_2 -0.5) + 6(x_3-0.5) - 5(x_4^2 -0.3))
\] for Gaussian and binomial outcomes. Gaussian outcomes require
specification of dispersion, where we set the dispersion parameter to be
1. In this data generating process, we have \(x_1, x_2, x_3, x_4\) as
the active predictors, while the rest predictors are inactive,
i.e.~\(f_j(x_j) = 0\) for \(j = 4, \dots, p\). Another set of
independent sample of size \(n_{test}=1000\), is created following the
same data generating process, serving as the testing data. We generate
50 independent pairs of training and testing datasets to evaluate the
prediction and variable selection performance of the chosen models,
where training datasets are used to fit the models and testing datasets
are used to calculate metrics of interest. In addition, we consider the
data generating process where all functional forms of the predictors are
linear while keeping the rest of simulation parameters the same. This
additional set of linear simulations is designed to investigate the
flexibility of the proposed model when nonlinear assumptions are not
met.

To evaluate the predictive performance of the models, the statistics,
\(R^2\) for Gaussian model and AUC for binomial model calculated based
on the testing datasets, are averaged across 50 simulations. To evaluate
the variable selection performance of the models, we record the set of
variables each method selects and calculate the averaged positive
predictive value (precision), true positive rate (recall), and Matthews
correlation coefficient (MCC), \begin{align*}
&\text{precision} = \frac{TP}{PP}\\
&\text{recall} = \frac{TP}{TP+FP}\\
&\text{MCC} = \frac{TP\times TN - FP \times FN}{\sqrt{(TP+FP)(TP+FN)(TN+FP)(TN+FN)}},
\end{align*} where TP, TN, FP, FN, and PP are true positives, true
negatives, false positives, false negatives, and predicted positives
respectively. For the methods that don't automatically achieve variable
selection, we set the alpha level at 0.05 for \textit{mgcv} that relies
on hypothesis testing, and a soft-threshold at 0.5 for spikeSlabGAM
given the marginal inclusion probabilities. For the two methods, BHAM
and spikeSlabGAM, that are capable of bi-level selection, we record the
probability that the linear and nonlinear components of each predictor
are selected in the models.

\subsection{Simulation Results}
\subsubsection{Prediction Performance}

Among the set of simulations where the functional forms of the
predictors are nonlinear, the predictive performances have a consistent
pattern across the two distributions of outcomes. For simplicity, we use
Gaussian simulations to exemplify the improvement of BHAM and defer to
Tables \ref{tab:gaus} and \ref{tab:bin_auc} for detailed statistics. The
proposed model, BHAM, predicts as good as, if not better than, other
high dimensional additive models. Specifically, BHAM shows great
improvement over COSSO methods, resulting in a median (interquartile
range, IRT) 31\% (131\%) and 20\% (129\%) improvement over COSSO and
adaptive COSSO in \(R^2\) statistics respectively. The improvement over
the spikeSlabGAM model is moderate, resulting in a median (IRT) 6\%
(10\%) improvement in \(R^2\). When comparing to SB-GAM, BHAM performs
better (median (IRT) 13\% (8\%) improvement in R2) in lower dimensional
cases (\(p=4,10\)), and equally good or slightly worse (median (IRT) 1\%
(9\%) improvement in R2: ) in high-dimensional cases
(\(p=50,100, 200\)). As previously hypothesized, the linear LASSO model
predicts less accurate than other flexible models across all scenarios;
mgcv performs extremely well in low-dimensional cases (\(p = 4, 10\)),
and deteriorates as the dimensionality increases until not applicable.
To note, mgcv fits models but fails to converge within the default
number of iterations when the sample size approaches the number of
coefficients to estimate (\(p=50\)), which leads to bad performance.
Even though SB-GAM has slight prediction advantage over the proposed
model in high-dimensional situations, the BHAM has an extreme
computational advantage over SB-GAM, resulting median (IRT) 64\% (39\%)
reduction in computation time (measured in seconds) for Gaussian
simulations, without sacrificing much of the prediction accuracy (see
Table \ref{tab:time_sim}).

We also examine the prediction performance when the functional form of
predictors is linear, see Supporting Information Table S1, and S2. The
proposed model, BHAM, has a similar performance as the linear LASSO
model regardless of the distribution. This observation demonstrates that
BHAM is a flexible model, and has good prediction performance regardless
of the underlying functional form of predictors. spikeSlabGAM has a
similar prediction performance to BHAM. Surprisingly, SB-GAM doesn't
perform well in high-dimensional Gaussian outcome scenarios.

\subsubsection{Variable Selection Performance}

Among the set of simulations where the functional forms of the
predictors are nonlinear, the proposed model, BHAM, has a consistent
performance across different dimensions and distribution settings (See
Table \ref{tab:sim_gaus_var_select} for Gaussian outcomes, and Support
Information Table S3 for binomial outcomes): being conservative. The
symptoms of conservative variable selection are high precision and low
recall, where high precision means that among all the selected
variables, high percentage of them are true signals; low recall means
that the model selected a small subset among all the active predictions.
In other words, BHAM tends to select a smaller set of variables that are
truly effective to the outcome. We want to note, the variable selection
performance of BHAM is plummeted and not optimized when \(p=200\). Upon
further investigation, we discover it's because the generic sequence of
\(s_0\) used to tune the model doesn't contain the optimal value.
Overall, among all the models examined, SB-GAM has the best performance,
both high precision and high recall, and yields a high MCC. The
performance of another Bayesian model, spikeSlabGAM deteriorates as the
sparsity grows, particularly when (\(p>50\)), or for binomial outcomes.
The variable selection performance for linear simulations matches with
prediction performance: BHAM performs great among the Gaussian
scenarios, while the performance of SB-GAM deteriorates.

Among the high-dimensional methods of comparison, there are two methods
that are capable to achieve bi-level selection, the proposed BHAM and
spikeSlabGAM. Among the linear simulations, both methods can accurately
select the linear components and have a drastically lowered probability,
close to 0, to include the nonlinear component, as anticipated.
Specifically, spikeSlabGAM have a smaller probability to include the
nonlinear component in the model than BHAM. However, this advantage of
spikeSlabGAM over BHAM is less obvious among the nonlinear simulations:
spikeSlabGAM performs better than BHAM when selecting components of the
functional forms that include only linear or nonlinear component,
e.g.~functional forms for \(x_3\) and \(x_4\). However, spikeSlabGAM
inclines to ignore variables that have more complex function forms,
e.g.~function forms for \(x_1\) and \(x_2\). In contrast, BHAM is more
likely to include them in the model. This trade-off is determined by the
assumption implicitly reflected via the prior hierarchy. We defer an
in-depth discussion to Section \ref{sec:concl}.

\section{Metabolomics Data Analysis}
\label{sec:real_data}

In this section, we apply the proposed model BHAM to analyze two
published metabolomics datasets where the outcomes are binary and
continuous respectively. We demonstrate the improved prediction
performance compared to the other Bayesian hierarchical additive model,
SB-GAM \citep{Bai2021}, while being computationally efficient (see Table
\ref{tab:time_real_data}). BHAM requires roughly 10\% of the computation
time of SB-GAM to fit models.

\subsection{Emory Cardiovascular Biobank}
\label{sec:ECB}

\begin{figure}[h] 
\includegraphics{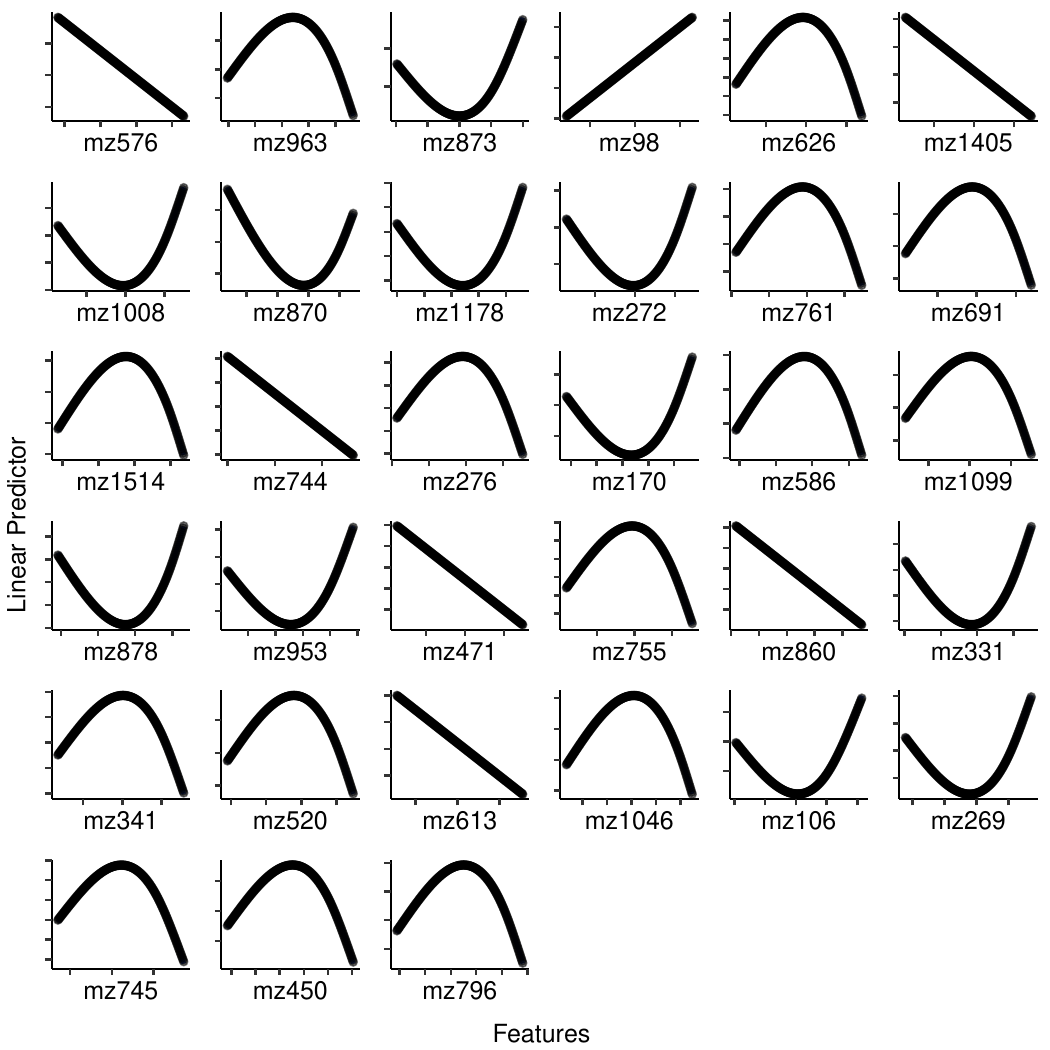}
\caption{Plots of the functions for the 33 metablites selected by BHAM in the Emory Cardiovascular Biobank data analysis}
\label{fig:ECB_fig}
\end{figure}

We use the proposed model BHAM to analyze a metabolic dataset from a
recently published research \citep{Mehta2020} studying plasma
metabolomic profile on the three-year all-cause mortality among patients
undergoing cardiac catheterization. The dataset is publicly available
via \textit{Dryad} \citep{Mehta2020_data}. It contains in total of 776
subjects from two cohorts. As there is a large number of non-overlapping
features among the two cohorts, we use the cohort with a larger sample
size (N=454). There are initially 6796 features in the dataset, which is
too large to be practically meaningful to analyze. Hence, we choose the
top 200 features with the largest variance. We use 5-knot spline
additive models for binary outcome using two different models, the
proposed BHAM and the SB-GAM. 10-Fold CV are used to choose the optimal
tuning parameters of each framework with respect to the default
selection criterion implemented in the software. Out-of-bag samples are
used for prediction performance evaluation, where deviance, AUC, Brier
score, defined as
\(\frac{1}{n}\sum\limits^{n}_{i=1}(y_i - \hat y_i)^2\), and
misclassification error, defined as
\(\frac{1}{n}\sum\limits^{n}_{i=1}I(|y_i - \hat y_i|>0.5)\) are
calculated. BHAM obtains superior AUC, Brier score, and
misclassification error in the out-of-bag samples compared to SB-GAM
(see Table \ref{tab:ECB_res}). We plot the 33 features included in the
final BHAM model in Figure \ref{fig:ECB_fig}.

\subsection{Weight Loss Maintenance Cohort}

We use the proposed model BHAM to analyze metabolomics data from a
recently published study \citep{Bihlmeyer2021} on the association
between metabolic biomarkers and weight loss, where the dataset is
publicly available \citep{Bihlmeyer2021_data}. In this analysis, we
primarily focus on the analysis of one of the three studies included,
weight loss maintenance cohort \citep{Svetkey2008}, due to the
drastically different intervention effects. In the dataset, 765
metabolites in baseline plasma collected were profiled using liquid
chromatography mass spectrometry. Quality control and natural log
transformation were previously performed and documented by the study
publishing team \cite{Bihlmeyer2021}. The outcome of interest is
standardized percent change in insulin resistance, and hence modeled
using a Gaussian model. After removing missing datapoints and addressing
outliers in the data, there are \(p\)=237 features remaining in the
analysis. 5-Knot spline additive models for the Gaussian outcome are
constructed using two different models, the proposed BHAM and the
SB-GAM. 10-Fold CV are used to choose the optimal tuning parameters of
each framework with respect to the default selection criterion
implemented in the software. Out-of-bag samples are used for prediction
performance evaluation, where deviance, \(R^2\), mean squared error
(MSE) defined as \(\frac{1}{n}\sum\limits^{n}_{i=1}(y_i - \hat y_i)^2\),
and mean absolute error (MAE) defined as
\(\frac{1}{n}\sum\limits^{n}_{i=1}|y_i - \hat y_i|\) are calculated.
BHAM obtains superior \(R^2\), MSE, and MAE in the out-of-bag samples
compared to SB-GAM (see Table \ref{tab:WLM_res}).

\section{Discussion}
\label{sec:concl}

In the paper, we described a novel high-dimensional generalized additive
model with Bayesian hierarchical prior for the purpose of predictive
modeling. In particular, we introduce a two-part spike-and-slab LASSO
prior for reparameterized smooth functions and derive a scalable EM-CD
algorithm for model fitting. The proposed model provides a new angle to
address the excess shrinkage of smooth functions that is commonly
vulnerable to previous regularized high-dimensional GAMs, and hence
improves the predictive performance. Th EM-CD algorithm, extended from
previous spike-and-slab LASSO models, provides a computationally
efficient alternative to the computational prohibitive MCMC algorithms,
enhancing the scalability of spike-and-slab models. In addition, the
two-part prior motivates the bi-level selection of predictors, selection
of linear and nonlinear components. In the simulation study and
real-data analyses, the proposed model demonstrates improvement in
prediction and computational advantage when compared to the
state-of-the-art models. When serving the purpose of variable selection,
trade-offs exist among methods of comparison. We implement the proposed
model in an open-source R package \texttt{BHAM}, deposited at
\url{https://github.com/boyiguo1/BHAM}. To maximize the flexibility of
smooth function specification, we deploy the same programming grammar as
in the state-of-the-art package \texttt{mgcv}, in contrast to previous
tools where smooth functions are limited to the default ones. Ancillary
functions are provided for model specification in high-dimensional
settings, curve plotting and functional selection.

The proposed model shares many commonalities with the SB-GAM
\citep{Bai2021}, which is independently developed around the same time
as the proposed work. Both frameworks emphasize computational efficiency
by deploying group spike-and-slab LASSO type priors and
optimization-based scalable algorithms. Bai provides the theoretical
proof for the consistency of variable selection using group
spike-and-slab LASSO prior. The proposed model focuses on improving
prediction performance for high-dimensional GAM, with the capability of
bi-level selection. Moreover, the proposed model can easily generalize
to other families of priors and smooth functions if desired. Not focused
in this manuscript, the generalization is described in the Supporting
Information.

During designing and analyzing the simulation study, we made couple of
interesting observations. First of all, variable selection is a delicate
topic in the context of predictive modeling. When prediction performance
is used to tune a model, the model could possibly include noise
variables in models, for example LASSO and LASSO-based models.
\cite{Wu2019} Moreover, bi-level selection is a more complex problem
than variable selection. The complexity shows on the validity of the
effect hierarchy principle. While most functional forms follow that the
linear component exists in the nonlinear function, there are functions
that don't follow it, e.g.~\(x^2\). The proposed prior and spikeSlabGAM
employ different structures: the proposed prior imposes effect hierarchy
while spikeSlabGAM treats the selection of linear and nonlinear
components independent. The different prior setups lead to trade-offs
for the purpose of bi-level selection. We recommend to use more judgment
in bi-level selection, either relying on heuristic knowledge to choose
appropriate prior or exploring multiple models when heuristic knowledge
doesn't exist. Secondly, we find the performance of the proposed model
is more sensitive to the granularity of \(s_0\) sequence in the
high-dimensional settings than in the lower dimension settings. Even
though the current default sequence of \(s_0\) can result in reasonable
performance shown in the simulation studies, we recommend fine-tuning
the model with a granular sequence of \(s_0\) for performance
improvement.

Our future efforts are direct to uncertainty inference of the proposed
model, survival analysis and integrative analysis. Using EM-CD algorithm
to fit the proposed BHAM is incapable of conducting uncertainty
inference. We derive the EM-Iterative Weighted Least Square algorithm
(EM-IWLS, see the Supporting Information) as an alternative. Instead of
the Coordinate Descent algorithm, we use the Iterative Weighted Least
Square algorithm in the EM procedure. The EM-IWLS algorithm is
previously used to fit Bayesian high-dimensional generalized linear
models \cite{Yi2012}, and deliver estimates of the coefficient
variance-covariance matrix. Due to the space limit, technical details
will be explained in a future manuscript. While the proposed model
addresses a great deal of analytic problems, analyzing the time-to-event
outcome remains unsolved. A naive approach would be convert a
time-to-event outcome to a Poisson outcome following Whitehead
\citep{Whitehead1980}. However, it would be more efficient to directly
fit Cox models via penalized pseudo likelihood function
\citep{Simon2011}. Meanwhile, with the growing understanding of
biological structure within -omics field, it is appealing to integrate
external biology information in the modeling process. The main
motivation for integrative models is that biologically informed grouping
of weak effects increases the power of detecting true associations
between features and the outcome \citep{Peterson2016}, and stabilizes
the analysis results for reproducibility purposes. Such integration can
be achieved by setting up a structural hyperprior on the inclusion
indicator of the smooth function null space \(\boldsymbol{\gamma}^0\). A
similar strategy has been used in Ferrari and Dunson
\citep{Ferrari2020}.

\clearpage

\bibliography{bibfile.bib}

\begin{thebibliography}{10}
\providecommand \doibase [0]{http://dx.doi.org/}%

\bibitem{Mallick2013}
Mallick H, Yi N. {Bayesian Methods for High Dimensional Linear Models}. {\it
  Journal of Biometrics {\&} Biostatistics} 2013(205)\string: 1--27.
\newblock \href {\doibase 10.4172/2155-6180.S1-005} {doi:
  10.4172/2155-6180.S1-005}

\bibitem{Breiman2001}
Breiman L. Statistical modeling: The two cultures (with comments and a
  rejoinder by the author). {\it Statistical science} 2001\string;
  16(3)\string: 199--231.

\bibitem{Hastie1987}
Hastie T, Tibshirani R. {Generalized additive models: Some applications}. {\it
  Journal of the American Statistical Association} 1987\string; 82(398)\string:
  371--386.
\newblock \href {\doibase 10.1080/01621459.1987.10478440} {doi:
  10.1080/01621459.1987.10478440}

\bibitem{Ravikumar2009}
Ravikumar P, Lafferty J, Liu H, Wasserman L. {Sparse additive models}. {\it
  Journal of the Royal Statistical Society: Series B (Statistical Methodology)}
  2009\string; 71(5)\string: 1009--1030.
\newblock \href {\doibase 10.1111/j.1467-9868.2009.00718.x} {doi:
  10.1111/j.1467-9868.2009.00718.x}

\bibitem{Yuan2006}
Yuan M, Lin Y. Model selection and estimation in regression with grouped
  variables. {\it Journal of the Royal Statistical Society: Series B
  (Statistical Methodology)} 2006\string; 68(1)\string: 49--67.

\bibitem{Huang2010}
Huang J, Horowitz JL, Wei F. Variable selection in nonparametric additive
  models. {\it Annals of statistics} 2010\string; 38(4)\string: 2282.

\bibitem{Wang2007}
Wang L, Chen G, Li H. Group SCAD regression analysis for microarray time course
  gene expression data. {\it Bioinformatics} 2007\string; 23(12)\string:
  1486--1494.

\bibitem{Xue2009}
Xue L. Consistent variable selection in additive models. {\it Statistica
  Sinica} 2009\string: 1281--1296.

\bibitem{Fan2001}
Fan J, Li R. Variable selection via nonconcave penalized likelihood and its
  oracle properties. {\it Journal of the American statistical Association}
  2001\string; 96(456)\string: 1348--1360.

\bibitem{Xu2015}
Xu X, Ghosh M, others . Bayesian variable selection and estimation for group
  lasso. {\it Bayesian Analysis} 2015\string; 10(4)\string: 909--936.

\bibitem{Yang2020}
Yang X, Narisetty NN, others . Consistent group selection with Bayesian high
  dimensional modeling. {\it Bayesian Analysis} 2020\string; 15(3)\string:
  909--935.

\bibitem{Bai2020}
Bai R, Moran GE, Antonelli JL, Chen Y, Boland MR. Spike-and-slab group lassos
  for grouped regression and sparse generalized additive models. {\it Journal
  of the American Statistical Association} 2020\string: 1--14.

\bibitem{Bai2021}
Bai R. Spike-and-Slab Group Lasso for Consistent Estimation and Variable
  Selection in Non-Gaussian Generalized Additive Models. {\it
  arXiv:2007.07021v5.} Preprint posted online June 5, 2021.
  https://arxiv.org/abs/2007.07021.

\bibitem{Scheipl2013}
Scheipl F, Kneib T, Fahrmeir L. Penalized likelihood and Bayesian function
  selection in regression models. {\it AStA Advances in Statistical Analysis}
  2013\string; 97(4)\string: 349--385.

\bibitem{Scheipl2012}
Scheipl F, Fahrmeir L, Kneib T. {Spike-and-slab priors for function selection
  in structured additive regression models}. {\it Journal of the American
  Statistical Association} 2012\string; 107(500)\string: 1518--1532.
\newblock \href {\doibase 10.1080/01621459.2012.737742} {doi:
  10.1080/01621459.2012.737742}

\bibitem{Wood2017}
Wood SN. {\it {Generalized additive models: An introduction with R, second
  edition}} .
\newblock 2017

\bibitem{Meier2009}
Meier L, {Van De Geer} S, B{\"{u}}hlmann P. {High-dimensional additive
  modeling}. {\it Annals of Statistics} 2009\string; 37(6 B)\string:
  3779--3821.
\newblock \href {\doibase 10.1214/09-AOS692} {doi: 10.1214/09-AOS692}

\bibitem{Marra2011}
Marra G, Wood SN. Practical variable selection for generalized additive models.
  {\it Computational Statistics \& Data Analysis} 2011\string; 55(7)\string:
  2372--2387.

\bibitem{Rockova2018b}
Ro{\v{c}}kov{\'{a}} V. {Bayesian estimation of sparse signals with a continuous
  spike-and-slab prior}. {\it The Annals of Statistics} 2018\string;
  46(1)\string: 401--437.
\newblock \href {\doibase 10.1214/17-AOS1554} {doi: 10.1214/17-AOS1554}

\bibitem{Rockova2018}
Ro{\v{c}}kov{\'{a}} V, George EI. {The Spike-and-Slab LASSO}. {\it Journal of
  the American Statistical Association} 2018\string; 113(521)\string: 431--444.
\newblock \href {\doibase 10.1080/01621459.2016.1260469} {doi:
  10.1080/01621459.2016.1260469}

\bibitem{Bai2021Review}
Bai R, Rockova V, George EI. Spike-and-Slab Meets LASSO: A Review of the
  Spike-and-Slab LASSO. {\it arXiv:2010.06451.} Preprint posted online July 1,
  2021. https://arxiv.org/abs/2010.06451.

\bibitem{Tang2018}
Tang Z, Shen Y, Li Y, et al. {Group spike-And-slab lasso generalized linear
  models for disease prediction and associated genes detection by incorporating
  pathway information}. {\it Bioinformatics} 2018\string; 34(6)\string:
  901--910.
\newblock \href {\doibase 10.1093/bioinformatics/btx684} {doi:
  10.1093/bioinformatics/btx684}

\bibitem{Tang2019}
Tang Z, Lei S, Zhang X, et al. {Gsslasso Cox: A Bayesian hierarchical model for
  predicting survival and detecting associated genes by incorporating pathway
  information}. {\it BMC Bioinformatics} 2019\string; 20(1)\string: 1--15.
\newblock \href {\doibase 10.1186/s12859-019-2656-1} {doi:
  10.1186/s12859-019-2656-1}

\bibitem{chipman2006prior}
Chipman H. Prior distributions for Bayesian analysis of screening experiments.
  In: Springer.  2006 (pp. 236--267).

\bibitem{George1997}
George EI, McCulloch RE. {Approaches for Bayesian variable selection.}. {\it
  Statistica Sinica} 1997\string; 7(2)\string: 339--373.

\bibitem{Rockova2014a}
Ro{\v{c}}kov{\'{a}} V, George EI. {EMVS: The EM approach to Bayesian variable
  selection}. {\it Journal of the American Statistical Association}
  2014\string; 109(506)\string: 828--846.
\newblock \href {\doibase 10.1080/01621459.2013.869223} {doi:
  10.1080/01621459.2013.869223}

\bibitem{Tang2017a}
Tang Z, Shen Y, Zhang X, Yi N. {The spike-and-slab lasso generalized linear
  models for prediction and associated genes detection}. {\it Genetics}
  2017\string; 205(1)\string: 77--88.
\newblock \href {\doibase 10.1534/genetics.116.192195} {doi:
  10.1534/genetics.116.192195}

\bibitem{Tang2017}
Tang Z, Shen Y, Zhang X, Yi N. {The spike-and-slab lasso Cox model for survival
  prediction and associated genes detection}. {\it Bioinformatics} 2017\string;
  33(18)\string: 2799--2807.
\newblock \href {\doibase 10.1093/bioinformatics/btx300} {doi:
  10.1093/bioinformatics/btx300}

\bibitem{Friedman2010}
Friedman J, Hastie T, Tibshirani R. Regularization paths for generalized linear
  models via coordinate descent. {\it Journal of statistical software}
  2010\string; 33(1)\string: 1.

\bibitem{Zhang2006GAM}
Zhang HH, Lin Y. Component selection and smoothing for nonparametric regression
  in exponential families. {\it Statistica Sinica} 2006\string: 1021--1041.

\bibitem{Storlie2011}
Storlie CB, Bondell HD, Reich BJ, Zhang HH. Surface estimation, variable
  selection, and the nonparametric oracle property. {\it Statistica Sinica}
  2011\string; 21(2)\string: 679.

\bibitem{Wood2011}
Wood SN. Fast stable restricted maximum likelihood and marginal likelihood
  estimation of semiparametric generalized linear models. {\it Journal of the
  Royal Statistical Society (B)} 2011\string; 73(1)\string: 3-36.

\bibitem{R}
{R Core Team} . R: A Language and Environment for Statistical Computing. {\it R
  Foundation for Statistical Computing} 2021. https://www.R-project.org/.

\bibitem{Mehta2020}
Mehta A, Liu C, Nayak A, et al. Untargeted high-resolution plasma metabolomic
  profiling predicts outcomes in patients with coronary artery disease. {\it
  PloS one} 2020\string; 15(8)\string: e0237579.

\bibitem{Mehta2020_data}
Mehta A, Liu C, Uppal K, Quyyumi A. Data from: Metabolomics - Emory
  Cardiovascular Biobank. {\it Dryad, Dataset} Retrived online August 17, 2021.
  https://doi.org/10.5061/dryad.866t1g1mt.

\bibitem{Bihlmeyer2021}
Bihlmeyer NA, Kwee LC, Clish CB, et al. Metabolomic profiling identifies
  complex lipid species and amino acid analogues associated with response to
  weight loss interventions. {\it Plos one} 2021\string; 16(5)\string:
  e0240764.

\bibitem{Bihlmeyer2021_data}
Bihlmeyer NA, Kwee LC, Clish CB, et al. {Metabolomic profiling identifies
  complex lipid species and amino acid analogues associated with response to
  weight loss interventions}. {\it Zenodo} Retrived online August 18, 2021.
  https://doi.org/10.5281/zenodo.4767969.

\bibitem{Svetkey2008}
Svetkey LP, Stevens VJ, Brantley PJ, et al. Comparison of strategies for
  sustaining weight loss: the weight loss maintenance randomized controlled
  trial. {\it Jama} 2008\string; 299(10)\string: 1139--1148.

\bibitem{Wu2019}
Wu J, Witten D. {Flexible and Interpretable Models for Survival Data}. {\it
  Journal of Computational and Graphical Statistics} 2019\string; 28(4)\string:
  954--966.
\newblock \href {\doibase 10.1080/10618600.2019.1592758} {doi:
  10.1080/10618600.2019.1592758}

\bibitem{Yi2012}
Yi N, Ma S. {Hierarchical Shrinkage Priors and Model Fitting for
  High-dimensional Generalized Linear Models}. {\it Statistical Applications in
  Genetics and Molecular Biology} 2012\string; 11(6).
\newblock \href {\doibase 10.1515/1544-6115.1803} {doi: 10.1515/1544-6115.1803}

\bibitem{Whitehead1980}
Whitehead J. Fitting Cox's regression model to survival data using GLIM. {\it
  Journal of the Royal Statistical Society: Series C (Applied Statistics)}
  1980\string; 29(3)\string: 268--275.

\bibitem{Simon2011}
Simon N, Friedman J, Hastie T, Tibshirani R. Regularization paths for Cox’s
  proportional hazards model via coordinate descent. {\it Journal of
  statistical software} 2011\string; 39(5)\string: 1.

\bibitem{Peterson2016}
Peterson CB, Stingo FC, Vannucci M. Joint Bayesian variable and graph selection
  for regression models with network-structured predictors. {\it Statistics in
  medicine} 2016\string; 35(7)\string: 1017--1031.

\bibitem{Ferrari2020}
Ferrari F, Dunson DB. Identifying main effects and interactions among exposures
  using Gaussian processes. {\it The Annals of Applied Statistics} 2020\string;
  14(4)\string: 1743--1758.

\end{thebibliography}

\clearpage

\clearpage
\begin{table}[ht]
\centering
\begin{tabular}{cccccccc}
  \hline
P & mgcv & LASSO & COSSO & Adaptive COSSO & BHAM & SB-GAM & spikeSlabGAM \\ 
  \hline
  4 & 0.90 (0.01) & 0.33 (0.01) & 0.71 (0.13) & 0.72 (0.11) & 0.90 (0.01) & 0.79 (0.04) & 0.80 (0.00) \\ 
   10 & 0.90 (0.01) & 0.33 (0.01) & 0.66 (0.21) & 0.77 (0.02) & 0.89 (0.01) & 0.79 (0.04) & 0.79 (0.00) \\ 
   50 & 0.86 (0.02) & 0.32 (0.01) & 0.46 (0.19) & 0.57 (0.18) & 0.80 (0.02) & 0.78 (0.05) & 0.78 (0.01) \\ 
  100 & - & 0.32 (0.01) & 0.41 (0.23) & 0.48 (0.25) & 0.79 (0.01) & 0.79 (0.05) & 0.77 (0.01) \\ 
  200 & - & 0.32 (0.01) & 0.39 (0.19) & 0.40 (0.17) & 0.79 (0.01) & 0.78 (0.04) & 0.75 (0.01) \\ 
   \hline
\end{tabular}
\caption{The average and standard deviation of the out-of-sample $R^2$ measure for
    Gaussian outcomes over 50 iterations. The models of comparison include the proposed
    Bayesian hierarchical additive model (BHAM), linear LASSO model (LASSO), component
    selection and smoothing operator (COSSO), adaptive COSSO, mgcv, sparse Bayesian
    generalized additive model (SB-GAM), and spikeSlabGAM model. mgcv doesn't provide estimation
    when number of parameters exceeds sample size i.e. p = 100, 200.} 
\label{tab:gaus}
\end{table}

\clearpage
\begin{table}[ht]
\centering
\begin{tabular}{cccccccc}
  \hline
P & mgcv & LASSO & COSSO & Adaptive COSSO & BHAM & SB-GAM & spikeSlabGAM \\ 
  \hline
  4 & 0.94 (0.01) & 0.83 (0.01) & 0.90 (0.01) & 0.90 (0.01) & 0.92 (0.01) & 0.92 (0.01) & 0.90 (0.00) \\ 
   10 & 0.92 (0.03) & 0.83 (0.00) & 0.86 (0.04) & 0.86 (0.03) & 0.92 (0.01) & 0.92 (0.01) & 0.90 (0.00) \\ 
   50 & 0.76 (0.03) & 0.83 (0.01) & 0.83 (0.02) & 0.84 (0.02) & 0.90 (0.01) & 0.92 (0.01) & 0.89 (0.01) \\ 
  100 & - & 0.83 (0.01) & 0.83 (0.02) & 0.81 (0.09) & 0.90 (0.01) & 0.92 (0.01) & 0.88 (0.01) \\ 
  200 & - & 0.83 (0.01) & 0.81 (0.06) & 0.82 (0.05) & 0.88 (0.02) & 0.92 (0.01) & 0.87 (0.02) \\ 
   \hline
\end{tabular}
\caption{The average and standard deviation of the out-of-sample area under the curve measures
    for binomial outcomes over 50 iterations. The models of comparison include the proposed Bayesian
    hierarchical additive model (BHAM), linear LASSO model (LASSO), component selection and smoothing
    operator (COSSO), adaptive COSSO, mgcv, sparse Bayesian generalized additive model (SB-GAM), and
    spikeSlabGAM model. mgcv doesn't provide
    estimation when number of parameters exceeds sample size i.e. p = 100, 200.} 
\label{tab:bin_auc}
\end{table}

\clearpage
\begin{sidewaystable}[!h]
\centering
\begin{tabular}{lrllllll}
  \hline
Distribution & P & mgcv & COSSO & Adaptive COSSO & BHAM & SB-GAM & spikeSlabGAM \\ 
  \hline
Binomial &   4 & 0.18 (0.04) & 3.16 (1.39) & 5.51 (4.07) & 2.73 (0.22) & 347.17 (89.43) & 8.41 (0.91) \\ 
  Binomial &  10 & 3.46 (11.06) & 8.30 (1.70) & 10.66 (5.30) & 4.08 (0.29) & 539.05 (135.55) & 20.36 (2.16) \\ 
  Binomial &  50 & 660.31 (141.53) & 103.41 (20.00) & 118.82 (18.45) & 14.22 (0.58) & 1590.09 (142.19) & 236.73 (14.83) \\ 
  Binomial & 100 & - & 662.61 (125.00) & 672.65 (185.09) & 31.61 (2.56) & 2720.53 (250.43) & 967.97 (186.85) \\ 
  Binomial & 200 & - & 5325.66 (995.60) & 4963.93 (1482.11) & 82.17 (3.29) & 4788.76 (420.64) & 3371.88 (194.02) \\ 
  Gaussian &   4 & 0.05 (0.01) & 0.75 (0.09) & 0.75 (0.11) & 8.78 (1.57) & 38.82 (2.74) & 1.84 (0.18) \\ 
  Gaussian &  10 & 0.32 (0.39) & 3.42 (0.24) & 3.41 (0.23) & 20.77 (3.95) & 76.12 (5.55) & 5.93 (0.57) \\ 
  Gaussian &  50 & 72.03 (57.99) & 33.98 (2.88) & 34.35 (2.86) & 285.73 (12.53) & 374.76 (23.79) & 65.18 (8.12) \\ 
  Gaussian & 100 & - & 117.79 (3.33) & 119.63 (3.66) & 372.01 (56.92) & 640.44 (21.91) & 194.14 (8.09) \\ 
  Gaussian & 200 & - & 518.86 (40.78) & 524.76 (39.15) & 471.46 (72.23) & 1300.70 (72.74) & 738.52 (62.76) \\ 
   \hline
\end{tabular}
\caption{The average and standard deviation of computation time in seconds,
      including cross-validation and final model fitting, over 50 iterations. The models
      of comparison include the proposed Bayesian hierarchical additive model (BHAM), the linear LASSO model (LASSO),
      component selection and smoothing operator (COSSO), adaptive COSSO,
      mgcv, sparse Bayesian generalized additive model (SB-GAM), spikeSlabGAM. mgcv doesn't 
      provide estimation when number of parameters exceeds sample size i.e. p = 100, 200.} 
\label{tab:time_sim}
\end{sidewaystable}

\clearpage
\begin{table}[ht]
\centering
\begin{tabular}{cccccccc}
  \hline
P & Metric & LASSO & COSSO & Adaptive COSSO & BHAM & SB-GAM & spikeSlabGAM \\ 
  \hline
  4 & Precision & 1.00 & 1.00 & 1.00 & 1.00 & 1.00 & 1.00 \\ 
   10 & Precision & 0.76 & 0.84 & 0.93 & 0.62 & 0.86 & 1.00 \\ 
   50 & Precision & 0.48 & 0.70 & 0.69 & 0.88 & 0.75 & 1.00 \\ 
  100 & Precision & 0.43 & 0.61 & 0.59 & 0.99 & 0.79 & 0.99 \\ 
  200 & Precision & 0.36 & 0.61 & 0.47 & 0.28 & 0.75 & 0.99 \\ 
   \hline
  4 & Recall & 0.53 & 0.49 & 0.53 & 0.88 & 0.99 & 0.51 \\ 
   10 & Recall & 0.40 & 0.52 & 0.52 & 0.83 & 1.00 & 0.50 \\ 
   50 & Recall & 0.35 & 0.40 & 0.48 & 0.37 & 1.00 & 0.50 \\ 
  100 & Recall & 0.33 & 0.36 & 0.40 & 0.30 & 0.99 & 0.50 \\ 
  200 & Recall & 0.32 & 0.33 & 0.35 & 0.52 & 1.00 & 0.50 \\ 
   \hline
 10 & MCC & 0.32 & 0.49 & 0.57 & 0.46 & 0.86 & 0.61 \\ 
   50 & MCC & 0.31 & 0.47 & 0.53 & 0.50 & 0.83 & 0.69 \\ 
  100 & MCC & 0.32 & 0.41 & 0.45 & 0.53 & 0.87 & 0.70 \\ 
  200 & MCC & 0.28 & 0.41 & 0.38 & 0.36 & 0.85 & 0.70 \\ 
   \hline
\end{tabular}
\caption{The variable selection performance of Gaussian simulations,
                         measured by positive predictive value (precision), true positive rate (recall),
                         and Matthews correlation coefficient (MCC), for the high-dimensional methods
                         averaged over 50 iterations. The models of comparison include the proposed Bayesian
                         hierarchical additive model (BHAM), linear LASSO model (LASSO), component selection
                         and smoothing operator (COSSO), adaptive COSSO, sparse Bayesian generalized additive
                         model (SB-GAM), and spikeSlabGAM model. MCC is ill-defined when $p=4$ simulation
                         (no true negative), and hence omitted for all methods.} 
\label{tab:sim_gaus_var_select}
\end{table}

\clearpage

\clearpage

\providecommand{\docline}[3]{\noalign{\global\setlength{\arrayrulewidth}{#1}}\arrayrulecolor[HTML]{#2}\cline{#3}}

\setlength{\tabcolsep}{2pt}

\renewcommand*{\arraystretch}{1.5}

\begin{longtable}[c]{|p{0.75in}|p{0.75in}|p{0.75in}|p{0.75in}|p{0.75in}|p{0.75in}|p{0.75in}}

\caption{Model fitting time in seconds for two metabolomics data analyses, from Emory Cardiovascular Biobank (ECB) and Weight Loss Maintenance Cohort (WLM). It tabulates the computation time for cross-validation step (CV) and optimal model fitting step (Final), and total computation time (Total) for the proposed model BHAM and the model of comparison SB-GAM.
}\\

\hhline{>{\arrayrulecolor[HTML]{666666}\global\arrayrulewidth=2pt}->{\arrayrulecolor[HTML]{666666}\global\arrayrulewidth=2pt}->{\arrayrulecolor[HTML]{666666}\global\arrayrulewidth=2pt}->{\arrayrulecolor[HTML]{666666}\global\arrayrulewidth=2pt}->{\arrayrulecolor[HTML]{666666}\global\arrayrulewidth=2pt}->{\arrayrulecolor[HTML]{666666}\global\arrayrulewidth=2pt}->{\arrayrulecolor[HTML]{666666}\global\arrayrulewidth=2pt}-}

\multicolumn{1}{!{\color[HTML]{000000}\vrule width 0pt}>{\centering}p{\dimexpr 0.75in+0\tabcolsep+0\arrayrulewidth}}{} & \multicolumn{3}{!{\color[HTML]{000000}\vrule width 0pt}>{\centering}p{\dimexpr 2.25in+4\tabcolsep+2\arrayrulewidth}}{\fontsize{11}{11}\selectfont{\textcolor[HTML]{000000}{BHAM}}} & \multicolumn{3}{!{\color[HTML]{000000}\vrule width 0pt}>{\centering}p{\dimexpr 2.25in+4\tabcolsep+2\arrayrulewidth}!{\color[HTML]{000000}\vrule width 0pt}}{\fontsize{11}{11}\selectfont{\textcolor[HTML]{000000}{SB-GAM}}} \\

\hhline{~>{\arrayrulecolor[HTML]{666666}\global\arrayrulewidth=1pt}->{\arrayrulecolor[HTML]{666666}\global\arrayrulewidth=1pt}->{\arrayrulecolor[HTML]{666666}\global\arrayrulewidth=1pt}->{\arrayrulecolor[HTML]{666666}\global\arrayrulewidth=1pt}->{\arrayrulecolor[HTML]{666666}\global\arrayrulewidth=1pt}->{\arrayrulecolor[HTML]{666666}\global\arrayrulewidth=1pt}-}

\multicolumn{1}{!{\color[HTML]{000000}\vrule width 0pt}>{\centering}p{\dimexpr 0.75in+0\tabcolsep+0\arrayrulewidth}}{\multirow[c]{-2}{*}{\fontsize{11}{11}\selectfont{\textcolor[HTML]{000000}{Data}}}} & \multicolumn{1}{!{\color[HTML]{000000}\vrule width 0pt}>{\centering}p{\dimexpr 0.75in+0\tabcolsep+0\arrayrulewidth}}{\fontsize{11}{11}\selectfont{\textcolor[HTML]{000000}{CV}}} & \multicolumn{1}{!{\color[HTML]{000000}\vrule width 0pt}>{\centering}p{\dimexpr 0.75in+0\tabcolsep+0\arrayrulewidth}}{\fontsize{11}{11}\selectfont{\textcolor[HTML]{000000}{Final}}} & \multicolumn{1}{!{\color[HTML]{000000}\vrule width 0pt}>{\centering}p{\dimexpr 0.75in+0\tabcolsep+0\arrayrulewidth}}{\fontsize{11}{11}\selectfont{\textcolor[HTML]{000000}{Total}}} & \multicolumn{1}{!{\color[HTML]{000000}\vrule width 0pt}>{\centering}p{\dimexpr 0.75in+0\tabcolsep+0\arrayrulewidth}}{\fontsize{11}{11}\selectfont{\textcolor[HTML]{000000}{CV}}} & \multicolumn{1}{!{\color[HTML]{000000}\vrule width 0pt}>{\centering}p{\dimexpr 0.75in+0\tabcolsep+0\arrayrulewidth}}{\fontsize{11}{11}\selectfont{\textcolor[HTML]{000000}{Final}}} & \multicolumn{1}{!{\color[HTML]{000000}\vrule width 0pt}>{\centering}p{\dimexpr 0.75in+0\tabcolsep+0\arrayrulewidth}!{\color[HTML]{000000}\vrule width 0pt}}{\fontsize{11}{11}\selectfont{\textcolor[HTML]{000000}{Total}}} \\

\hhline{>{\arrayrulecolor[HTML]{666666}\global\arrayrulewidth=2pt}->{\arrayrulecolor[HTML]{666666}\global\arrayrulewidth=2pt}->{\arrayrulecolor[HTML]{666666}\global\arrayrulewidth=2pt}->{\arrayrulecolor[HTML]{666666}\global\arrayrulewidth=2pt}->{\arrayrulecolor[HTML]{666666}\global\arrayrulewidth=2pt}->{\arrayrulecolor[HTML]{666666}\global\arrayrulewidth=2pt}->{\arrayrulecolor[HTML]{666666}\global\arrayrulewidth=2pt}-}

\endfirsthead

\hhline{>{\arrayrulecolor[HTML]{666666}\global\arrayrulewidth=2pt}->{\arrayrulecolor[HTML]{666666}\global\arrayrulewidth=2pt}->{\arrayrulecolor[HTML]{666666}\global\arrayrulewidth=2pt}->{\arrayrulecolor[HTML]{666666}\global\arrayrulewidth=2pt}->{\arrayrulecolor[HTML]{666666}\global\arrayrulewidth=2pt}->{\arrayrulecolor[HTML]{666666}\global\arrayrulewidth=2pt}->{\arrayrulecolor[HTML]{666666}\global\arrayrulewidth=2pt}-}

\multicolumn{1}{!{\color[HTML]{000000}\vrule width 0pt}>{\centering}p{\dimexpr 0.75in+0\tabcolsep+0\arrayrulewidth}}{} & \multicolumn{3}{!{\color[HTML]{000000}\vrule width 0pt}>{\centering}p{\dimexpr 2.25in+4\tabcolsep+2\arrayrulewidth}}{\fontsize{11}{11}\selectfont{\textcolor[HTML]{000000}{BHAM}}} & \multicolumn{3}{!{\color[HTML]{000000}\vrule width 0pt}>{\centering}p{\dimexpr 2.25in+4\tabcolsep+2\arrayrulewidth}!{\color[HTML]{000000}\vrule width 0pt}}{\fontsize{11}{11}\selectfont{\textcolor[HTML]{000000}{SB-GAM}}} \\

\hhline{~>{\arrayrulecolor[HTML]{666666}\global\arrayrulewidth=1pt}->{\arrayrulecolor[HTML]{666666}\global\arrayrulewidth=1pt}->{\arrayrulecolor[HTML]{666666}\global\arrayrulewidth=1pt}->{\arrayrulecolor[HTML]{666666}\global\arrayrulewidth=1pt}->{\arrayrulecolor[HTML]{666666}\global\arrayrulewidth=1pt}->{\arrayrulecolor[HTML]{666666}\global\arrayrulewidth=1pt}-}

\multicolumn{1}{!{\color[HTML]{000000}\vrule width 0pt}>{\centering}p{\dimexpr 0.75in+0\tabcolsep+0\arrayrulewidth}}{\multirow[c]{-2}{*}{\fontsize{11}{11}\selectfont{\textcolor[HTML]{000000}{Data}}}} & \multicolumn{1}{!{\color[HTML]{000000}\vrule width 0pt}>{\centering}p{\dimexpr 0.75in+0\tabcolsep+0\arrayrulewidth}}{\fontsize{11}{11}\selectfont{\textcolor[HTML]{000000}{CV}}} & \multicolumn{1}{!{\color[HTML]{000000}\vrule width 0pt}>{\centering}p{\dimexpr 0.75in+0\tabcolsep+0\arrayrulewidth}}{\fontsize{11}{11}\selectfont{\textcolor[HTML]{000000}{Final}}} & \multicolumn{1}{!{\color[HTML]{000000}\vrule width 0pt}>{\centering}p{\dimexpr 0.75in+0\tabcolsep+0\arrayrulewidth}}{\fontsize{11}{11}\selectfont{\textcolor[HTML]{000000}{Total}}} & \multicolumn{1}{!{\color[HTML]{000000}\vrule width 0pt}>{\centering}p{\dimexpr 0.75in+0\tabcolsep+0\arrayrulewidth}}{\fontsize{11}{11}\selectfont{\textcolor[HTML]{000000}{CV}}} & \multicolumn{1}{!{\color[HTML]{000000}\vrule width 0pt}>{\centering}p{\dimexpr 0.75in+0\tabcolsep+0\arrayrulewidth}}{\fontsize{11}{11}\selectfont{\textcolor[HTML]{000000}{Final}}} & \multicolumn{1}{!{\color[HTML]{000000}\vrule width 0pt}>{\centering}p{\dimexpr 0.75in+0\tabcolsep+0\arrayrulewidth}!{\color[HTML]{000000}\vrule width 0pt}}{\fontsize{11}{11}\selectfont{\textcolor[HTML]{000000}{Total}}} \\

\hhline{>{\arrayrulecolor[HTML]{666666}\global\arrayrulewidth=2pt}->{\arrayrulecolor[HTML]{666666}\global\arrayrulewidth=2pt}->{\arrayrulecolor[HTML]{666666}\global\arrayrulewidth=2pt}->{\arrayrulecolor[HTML]{666666}\global\arrayrulewidth=2pt}->{\arrayrulecolor[HTML]{666666}\global\arrayrulewidth=2pt}->{\arrayrulecolor[HTML]{666666}\global\arrayrulewidth=2pt}->{\arrayrulecolor[HTML]{666666}\global\arrayrulewidth=2pt}-}\endhead

\multicolumn{1}{!{\color[HTML]{000000}\vrule width 0pt}>{\centering}p{\dimexpr 0.75in+0\tabcolsep+0\arrayrulewidth}}{\fontsize{11}{11}\selectfont{\textcolor[HTML]{000000}{ECB}}} & \multicolumn{1}{!{\color[HTML]{000000}\vrule width 0pt}>{\centering}p{\dimexpr 0.75in+0\tabcolsep+0\arrayrulewidth}}{\fontsize{11}{11}\selectfont{\textcolor[HTML]{000000}{100.8}}} & \multicolumn{1}{!{\color[HTML]{000000}\vrule width 0pt}>{\centering}p{\dimexpr 0.75in+0\tabcolsep+0\arrayrulewidth}}{\fontsize{11}{11}\selectfont{\textcolor[HTML]{000000}{3.5}}} & \multicolumn{1}{!{\color[HTML]{000000}\vrule width 0pt}>{\centering}p{\dimexpr 0.75in+0\tabcolsep+0\arrayrulewidth}}{\fontsize{11}{11}\selectfont{\textcolor[HTML]{000000}{104.4}}} & \multicolumn{1}{!{\color[HTML]{000000}\vrule width 0pt}>{\centering}p{\dimexpr 0.75in+0\tabcolsep+0\arrayrulewidth}}{\fontsize{11}{11}\selectfont{\textcolor[HTML]{000000}{2,659.0}}} & \multicolumn{1}{!{\color[HTML]{000000}\vrule width 0pt}>{\centering}p{\dimexpr 0.75in+0\tabcolsep+0\arrayrulewidth}}{\fontsize{11}{11}\selectfont{\textcolor[HTML]{000000}{20.9}}} & \multicolumn{1}{!{\color[HTML]{000000}\vrule width 0pt}>{\centering}p{\dimexpr 0.75in+0\tabcolsep+0\arrayrulewidth}!{\color[HTML]{000000}\vrule width 0pt}}{\fontsize{11}{11}\selectfont{\textcolor[HTML]{000000}{2,679.9}}} \\

\multicolumn{1}{!{\color[HTML]{000000}\vrule width 0pt}>{\centering}p{\dimexpr 0.75in+0\tabcolsep+0\arrayrulewidth}}{\fontsize{11}{11}\selectfont{\textcolor[HTML]{000000}{WLM}}} & \multicolumn{1}{!{\color[HTML]{000000}\vrule width 0pt}>{\centering}p{\dimexpr 0.75in+0\tabcolsep+0\arrayrulewidth}}{\fontsize{11}{11}\selectfont{\textcolor[HTML]{000000}{365.4}}} & \multicolumn{1}{!{\color[HTML]{000000}\vrule width 0pt}>{\centering}p{\dimexpr 0.75in+0\tabcolsep+0\arrayrulewidth}}{\fontsize{11}{11}\selectfont{\textcolor[HTML]{000000}{6.8}}} & \multicolumn{1}{!{\color[HTML]{000000}\vrule width 0pt}>{\centering}p{\dimexpr 0.75in+0\tabcolsep+0\arrayrulewidth}}{\fontsize{11}{11}\selectfont{\textcolor[HTML]{000000}{372.2}}} & \multicolumn{1}{!{\color[HTML]{000000}\vrule width 0pt}>{\centering}p{\dimexpr 0.75in+0\tabcolsep+0\arrayrulewidth}}{\fontsize{11}{11}\selectfont{\textcolor[HTML]{000000}{3,116.0}}} & \multicolumn{1}{!{\color[HTML]{000000}\vrule width 0pt}>{\centering}p{\dimexpr 0.75in+0\tabcolsep+0\arrayrulewidth}}{\fontsize{11}{11}\selectfont{\textcolor[HTML]{000000}{32.7}}} & \multicolumn{1}{!{\color[HTML]{000000}\vrule width 0pt}>{\centering}p{\dimexpr 0.75in+0\tabcolsep+0\arrayrulewidth}!{\color[HTML]{000000}\vrule width 0pt}}{\fontsize{11}{11}\selectfont{\textcolor[HTML]{000000}{3,148.7}}} \\

\hhline{>{\arrayrulecolor[HTML]{666666}\global\arrayrulewidth=2pt}->{\arrayrulecolor[HTML]{666666}\global\arrayrulewidth=2pt}->{\arrayrulecolor[HTML]{666666}\global\arrayrulewidth=2pt}->{\arrayrulecolor[HTML]{666666}\global\arrayrulewidth=2pt}->{\arrayrulecolor[HTML]{666666}\global\arrayrulewidth=2pt}->{\arrayrulecolor[HTML]{666666}\global\arrayrulewidth=2pt}->{\arrayrulecolor[HTML]{666666}\global\arrayrulewidth=2pt}-}

\end{longtable}

\label{tab:time_real_data}

\clearpage

\begin{table}[ht]
\centering
\begin{tabular}{lrrrr}
  \hline
Methods & Deviance & AUC & Brier & Misclass \\ 
  \hline
BHAM & 510.99 & 0.61 & 0.19 & 0.24 \\ 
  SB-GAM & 636.56 & 0.56 & 0.22 & 0.30 \\ 
   \hline
\end{tabular}
\caption{Prediction performance of BHAM fitted with Coordinate Descent algorithm (BHAM) and SB-GAM models for Emory Cardiovascular Biobank by 10-fold cross-validation, including deviance, area under the curve (AUC), Brier score, and misclassification error (Misclass) where class labels are defined using threshold = 0.5.} 
\label{tab:ECB_res}
\end{table}

\clearpage

\begin{table}[ht]
\centering
\begin{tabular}{lrrrr}
  \hline
Methods & Deviance & $R^2$ & MSE & MAE \\ 
  \hline
BHAM & 668.01 & 0.07 & 0.93 & 0.76 \\ 
  SB-GAM & 666.83 & 0.03 & 0.98 & 0.77 \\ 
   \hline
\end{tabular}
\caption{Prediction performance of BHAM fitted with Coordinate Descent algorithm (BHAM)  and SB-GAM models for Weight Loss Maintenance Cohort by 10-fold cross-validation, including deviance, $R^2$,  mean squared error (MSE), and mean absolute error (MAE).} 
\label{tab:WLM_res}
\end{table}

\clearpage

\end{document}